# Lipid thermodynamics : melting is molecular**

Sailaja Krishnamurty, Milen Stefanov, Tzonka Mineva, Sylvie Bégu, Jean Marie Devoisselle, Annick Goursot, Rui Zhu, Dennis R. Salahub*

[*] Dr. S. Krishnamurty, Dr. T. Mineva, Dr. S. Bégu, Prof. J.M. Devoisselle, Dr. A. Goursot
ICGM, UMR 5253 CNRS, Ecole de Chimie de Montpellier, 34296 Montpellier, France
Dipl.-Chem. M. Stefanov
Institute of Catalysis, Bulgarian Academy of Sciences, Sofia, Bulgaria
Dr. R. Zhu, Prof. D.R. Salahub
Department of Chemistry and Institute for Biocomplexity and Informatics, University of Calgary, 2500 University Drive NW, Calgary, Alberta, Canada T2N 1N4
Fax: (+1) 440....
E-mail: dsalahub@ucalgary.ca

[**] S.K. acknowledges the Institut Cartnot Montpellier CED2 for financial support.

Since the early fluid-mosaic model,[1] it has been recognized that lipid membranes are not only solvents for proteins but also have essential cellular functions. Their composition, the heterogeneity of their structures, and their dynamical properties have implications for these functions.[2] In particular, the fluid nature of the lipid membranes has a critical importance for the life of the cell. The lipid bilayer fluidity controls the lipid protein activity, but also acts as a sensor for temperature change.[3] The fluidity of the cell membranes, normally in the liquid crystalline state, is precisely regulated because lipids undergo phase changes in response to temperature. Phospholipids, according to their fatty acid compositions, have a specific main phase transition temperature, $T_m$, also called the melting temperature, between the gel and liquid crystalline states. The melting transition is accompanied by enthalpy and volume changes. Controlling the transition temperature of lipids is crucial for biological systems. For drug delivery lipid vesicles, this control would allow one to adjust the lipid fluidity according to processing, transport and release conditions. Many experimental methods have been used to quantify this particular property for bilayers of single lipids or mixtures of lipids and/or molecules, in particular cholesterol. In the last decade, more detailed information on the lipid melting processes has been provided by accurate microcalorimetric approaches and atomic force microscopy.[4, 5] Interestingly, the isobaric heat capacity and the volume expansion with temperature have a proportional relationship for a variety of lipids.[6-9] This correlation has been analyzed in terms of a proportional relationship between the enthalpy ($\Delta H$) and volume ($\Delta V$) changes in the melting transition, with similar proportional factors for different systems.[6] The explanation of this fact could be that enthalpy and volume changes are mainly due to the "melting" of the individual lipid molecules.[6] In other words, the main phase transition would be driven by intrinsic structural changes within the lipid molecules, whereas the changes of free volumes and intermolecular interactions could be considered as perturbations. We present here a computational experiment, which provides a fundamental basis for ascertaining this interpretation.

BOMD computations are performed for the dimyristoyl phosphatidylcholine (DMPC) molecule using Density Functional Theory augmented with a damped empirical dispersion correction (DFT-D) at temperatures below and above the experimental phase transition temperature of a hydrated DMPC bilayer, $T_m$, 295.1-298.1K.[10] Simulation temperatures (230, 250, 270, 282, 297, 301, 325 and 350 K) were chosen close enough to each other in order to have overlapping histograms of the potential-energy state density. This method gives clear evidence of solid-liquid transitions.[11, 12] Since the molecule is in a complete liquid state at 350 K (highly disordered), the corresponding geometric parameters are not reported in Figures 2 and 3. The total equilibrated simulation time amounts to 75 ps, with a minimum of 7 ps for the lowest temperature. An extensive preliminary study of the conformational space of DMPC (minimum energy structures at 0 K) has shown the existence of several isoenergetic ground-state conformers.[13] The initial structure for the present dynamics simulations corresponds to one of these conformers. Figure 1 illustrates snapshots of this conformer. Verification was made that the other conformers have similar dynamical behaviours.

The analysis of the atomic motions in the simulations at T ≤ 282 K reveals that the only substantial changes in torsion angles correspond to reorientational fluctuations of the tail methylene groups about the axes of the acyl chains. In other words, the planes of the alkyl carbon skeletons fluctuate between parallel and perpendicular. The structure in Figure 1a shows a typical perpendicular chain orientation, with the β chain rotated by about 90° with respect to the γ chain for T = 270 K. For more clarity, the inserts 1b and 1c illustrate parallel and perpendicular orientations, repectively, from a different point of view. The change from one tail orientation to the other is achieved through transitory intermediate conformations involving concomittant changes of torsion angles in the -C-O-CO-$CH_2$- glycerol-carbon frameworks.

Around T = 297 K, the equilibrated structural regime is much more slowly established than the equilibrated temperature and energy regimes. A typical snapshot at this temperature is given in the insert 1d. The structural fluctuations involve mostly perpendicular alkyl chains keeping an interchain distance comparable to that found for lower temperatures (Figure 1a, 2). However, one kink (alkane trans to gauche transformation) has been formed at the extremity of each chain.

Around T = 301 K, a kink occurs first in the middle of the β chain (C7), followed by a second kink at the extremity of the same chain. These structural changes imply that the interchain average distance increases with respect to 282 K (Figure 2).

Raising the temperature to 325 K and further on to 350 K leads to the sampling of more configurational states introducing more kinks and disorder in the chains. This disorder induces an enlarged interchain average distance but does not concern the other internal characteristic distances, which do not change significantly, in particular, the motions of the head remain correlated with those of the next glycerol (Figure 2).

As shown in Figure 2, the configurations sampled by the dynamics keep constant the most characteristic head and head ... glycerol interatomic distances over the whole range of temperatures.[13] The average distance Cβ...Cγ, which represents the average of all C... C distances between the β and γ chains is also not significantly affected up to 297 K, the reorientational fluctuations being included in the average value.

In order to demonstrate that the observed trans to gauche transformations starting at 297 K arise from a molecular phase transition, we have calculated the distance-fluctuation criterion for melting introduced by Berry et al. .[14] This parameter $\Delta_B$ is expressed as

$$\Delta_B = \frac{2}{N(N-1)} \sum_{i<j} \frac{\sqrt{\langle \Delta r_{ij}^2 \rangle - \langle r_{ij} \rangle^2}}{\langle r_{ij} \rangle}$$

with $r_{ij}$ defining interatomic distances and N the number of atoms. This criterion has been found particularly stable for finite disordered systems.[15] Its variation with temperature, illustrated in Figure 3, shows clearly that the melting starts at around 297 K.

For first order phase transitions, such as lipid melting, the internal energy of the system shows a discontinuity at the transition temperature. An energetic description of melting can thus be obtained using plots (histograms) of the density of the molecular states sampled by the dynamics at the studied temperatures.[11, 12] The histograms are presented in Figure 4. As shown on this Figure, the gel (230, 250, 270, 282 K) and liquid (297, 301, 325 and 350 K) states appear in two distinct groups, the liquid bands being broader and lower. The transition between these two regions is shown by the discontinuity occurring between 282 and 297 K.

A comparison between Figures 3 and 4 allows one to correlate the structural and energetic aspects of the melting process. Structurally, it appears that, at 297 K, the molecule has kept most of the gel characteristics, despite the formation of kinks at the extremity of the chains, because of the absence of a substantial disorder which would alter their relative positions. This structural disorder starts at 301 K with the formation of a kink in the middle of the alkyl chain and evolves at higher temperatures. Energetically, the 297 K histogram belongs clearly to the transition regime. One can thus conclude that melting begins at around

297 K or slightly below, without displaying liquid state characteristics of chain disorder, which start to be manifested at 301 K and higher temperatures. On this structural basis, a lower bound value of the melting enthalpy can be estimated as the change from 297 to 301 K of the weighted molecular state energies, leading to the value of 18.5 kJ/mol.

Experimentally, the main phase transition has indeed been attributed to the appearance of trans-gauche transformations in the alkyl chains.[16] Our simulated melting temperature is very close to the experimental value obtained for a fully hydrated bilayer, i.e. 295 ± 1.5 K, i.e. within the experimental error.[5, 6, 10] This result indicates that the DMPC main phase transition temperature is indeed driven by the intramolecular dynamical properties, as suggested from recent experiments.[6] Let us note that dehydration of a lecithin bilayer does not influence $T_m$ if the water concentration remains higher than about 15-20%.[17] Further dehydration leads to increasing $T_m$, which has been attributed to repulsive hydration forces between the lipid bilayers.[18] The melting enthalpy increases by about 2 kJ/mol for a water concentration below 20%.[17]

The observed melting transition of lipid bilayers is accompanied by enthalpy and volume increases. Melting transition enthalpy values of 22.7,[19] 23.4 ± 3.2,[6] 24.7[20] or 32.3 kJ/mol[9] have been obtained by calorimetry for DMPC unilamellar vesicles. Our calculated value is thus a very good estimate of the measured melting enthalpy. The evaluation of the volume expansion at Tm is in progress, using isodensity contours to define the molecular volume.

Finally, the analysis of the electronic energies of snapshots in the 297 and 300 K trajectories showed that the calculated melting enthalpy does not originate from non bonding intramolecular interactions, electrostatic or dispersion-like, but rather from the presence of gauche conformations in the chains and of the population of higher energy molecular states.

This computational experiment demonstrates that chain melting in lipids is a molecular process. BOMD is certainly the best method to reproduce such dynamics properties, since the electronic contributions to the various molecular structures sampled by the classical nuclear dynamics are described accurately.

Figure captions

Figure 1- a DMPC molecule (a) a typical snapshot of the 270 K dynamics; (b) illustration of parallel tails; (c) illustration of perpendicular tails; (d) a snapshot of the 297 K dynamics.

Figure 2- Selected interatomic distances averaged over the simulation time, as a function of temperature; the <C$\beta$- C$\gamma$> black line represents the averaged interchain distance between all the $\beta$ and $\gamma$ carbons.

Figure 3- The melting criterion $\Delta_B$ (normalized interatomic distance fluctuation) as a function of temperature.

Figure 4- The density of states histograms calculated for 230, 250, 270, 282, 297, 301, 325 and 350 K from left to right. The histograms are calculated on the last 7 ps trajectory at every temperature. E is the state energy and $E_0$ is the optimized electronic energy of the molecule (T = 0 K), taken as origin.

Figure 1

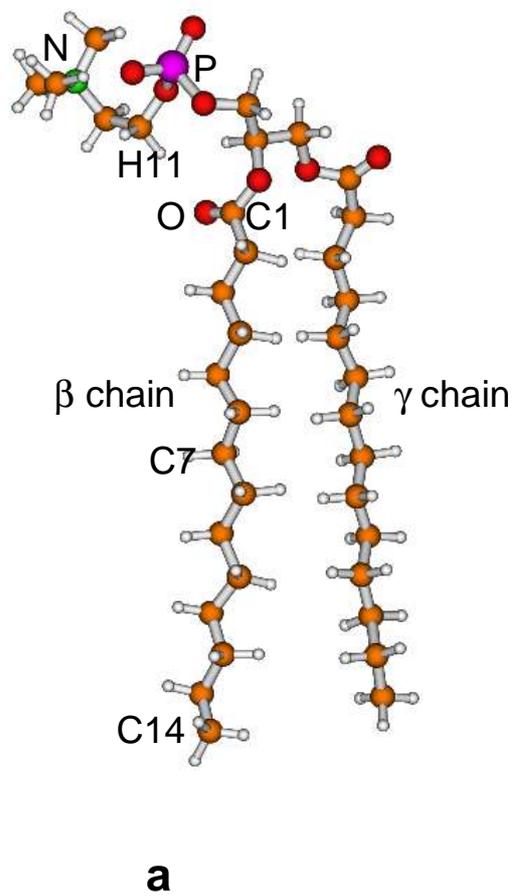
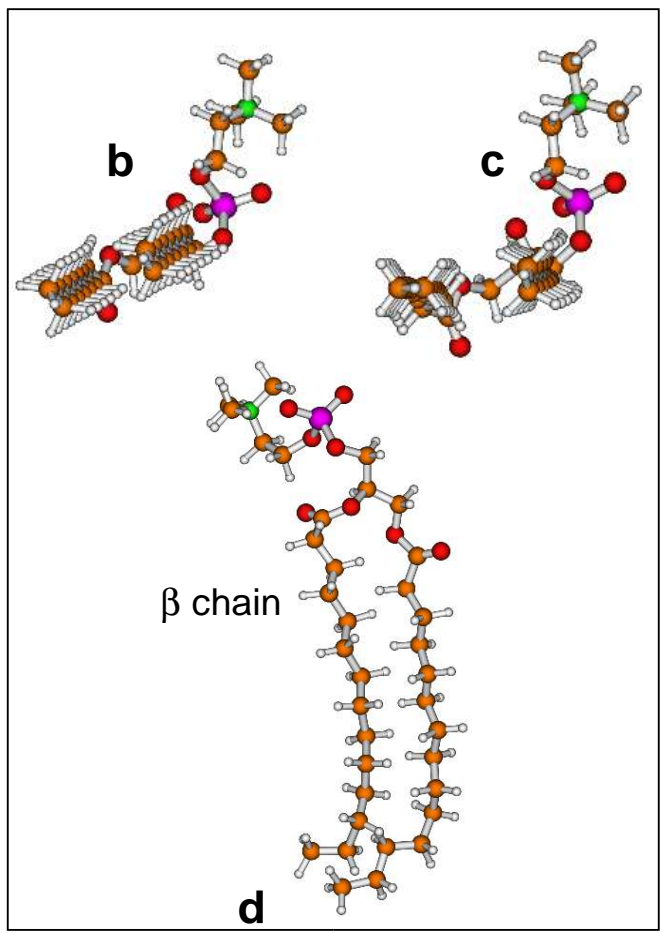

Figure 2

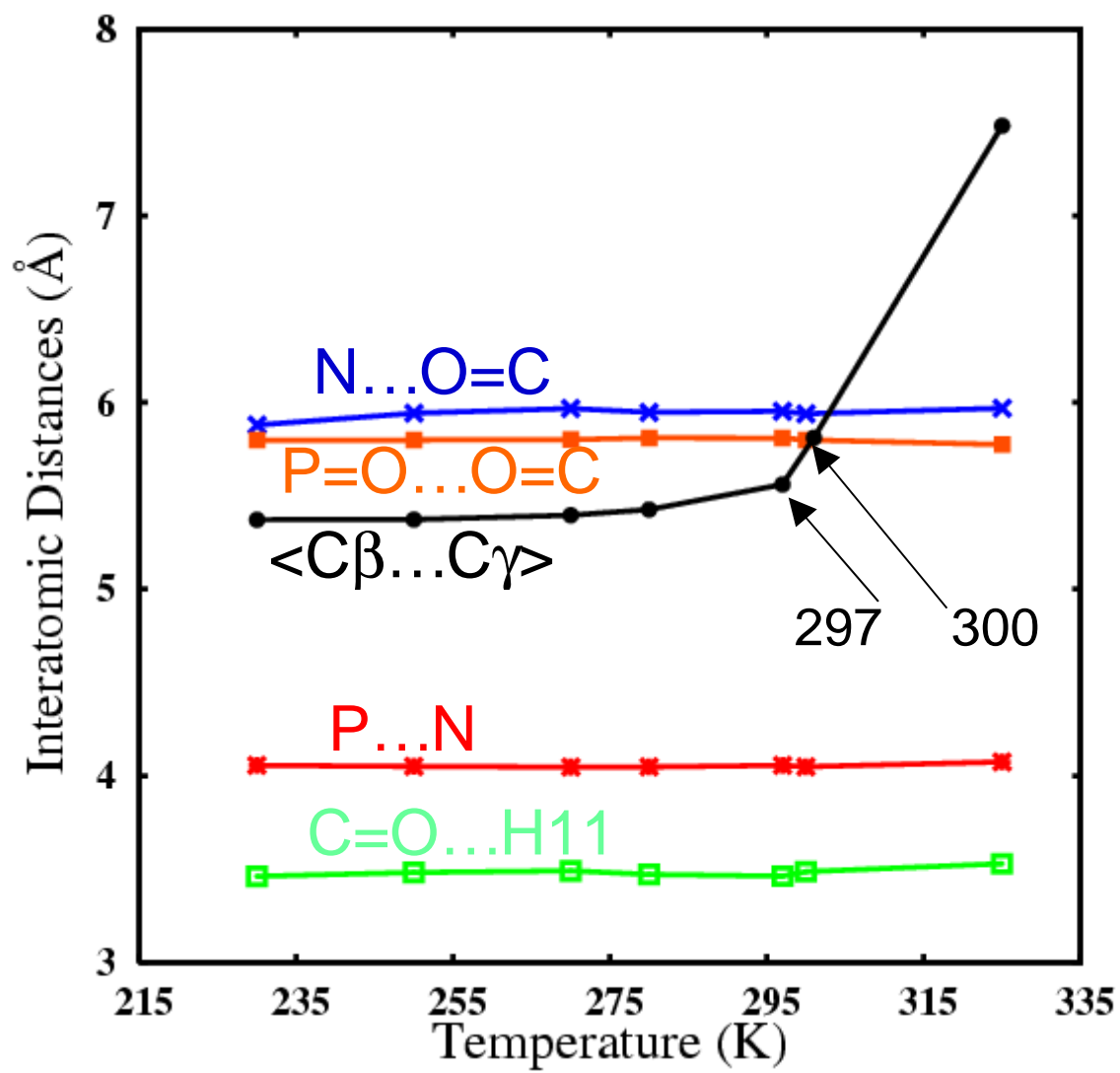

Figure 3

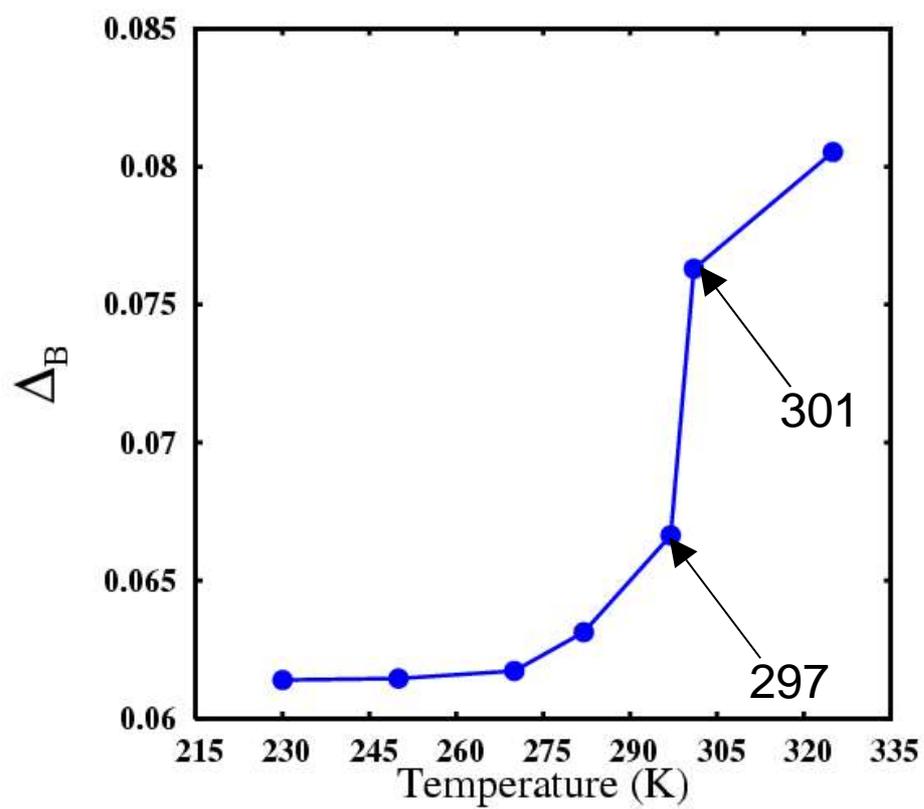

Figure 4

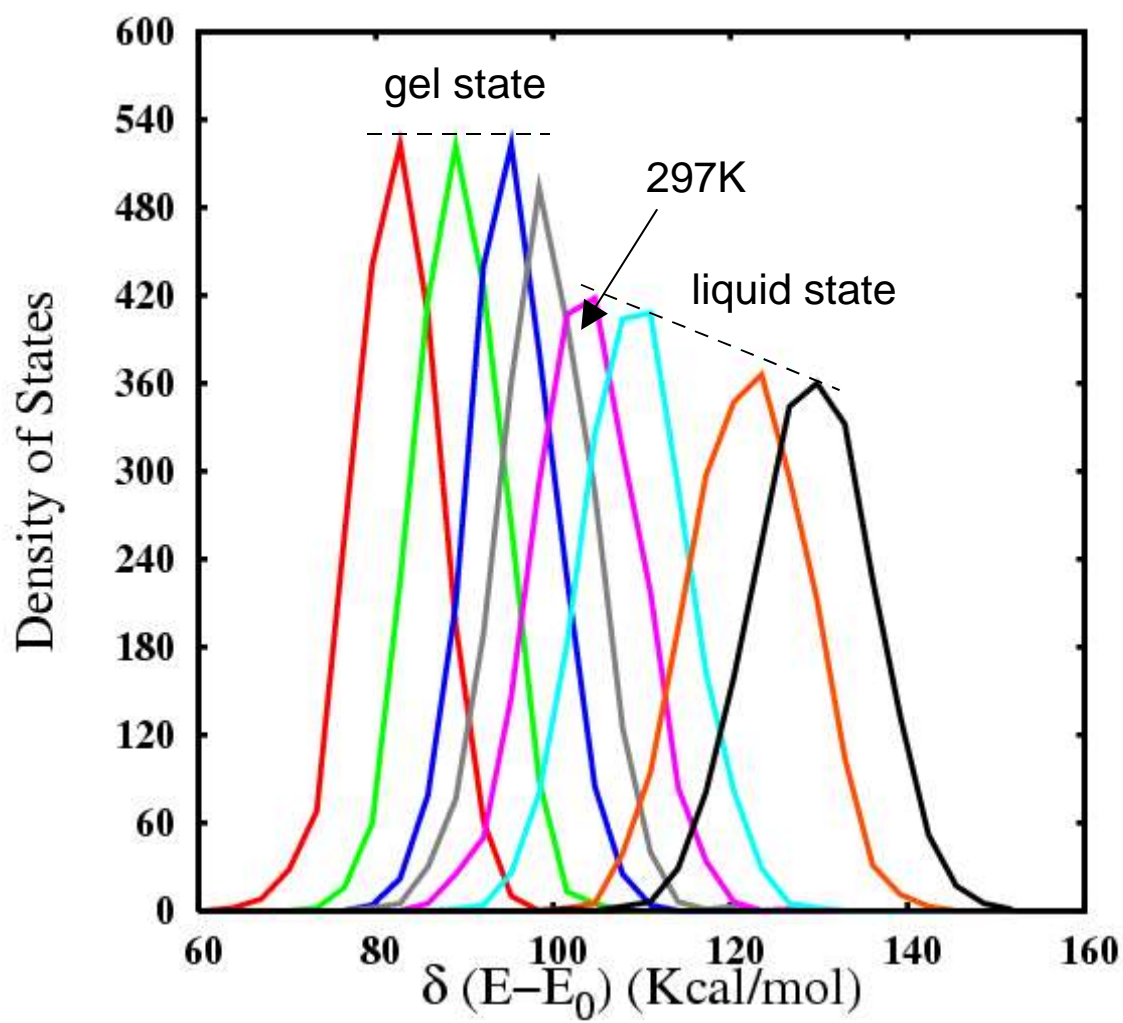